\documentclass{article}
\usepackage{spconf,amsmath,epsfig}
\usepackage{amsthm,amssymb}
\usepackage{graphicx}
\usepackage{svg}
\usepackage{hyperref}

\title{General Feature Extraction in SAR Target Classification: \\A Contrastive Learning Approach Across Sensor Types
}

\name{M. Muzeau$^{1, 2}$ \qquad
    J. Frontera-Pons$^{2}$ \qquad
    C. Ren$^{1}$ \qquad
    J.-P. Ovarlez$^{1, 2}$}
\address{ $^{1}$ SONDRA, CentraleSup\'elec, Universit\'e Paris-Saclay, 91192 Gif-sur-Yvette, France \\
    $^{2}$ DEMR, ONERA, Universit\'e Paris-Saclay, 91120 Palaiseau, France\\}

\begin{document}

\maketitle
\begin{abstract}
The increased availability of SAR data has raised a growing interest in applying deep learning algorithms. However, the limited availability of labeled data poses a significant challenge for supervised training. This article introduces a new method for classifying SAR data with minimal labeled images. The method is based on a feature extractor Vit trained with contrastive learning. It is trained on a dataset completely different from the one on which classification is made. The effectiveness of the method is assessed through 2D visualization using t-SNE for qualitative evaluation and k-NN classification with a small number of labeled data for quantitative evaluation. Notably, our results outperform a k-NN on data processed with PCA and a ResNet-34 specifically trained for the task, achieving a 95.9\% accuracy on the MSTAR dataset with just ten labeled images per class.
\end{abstract}
\section{Introduction}
Synthetic Aperture Radar (SAR) is a remote sensing technology that utilizes microwave signals to capture images of the Earth's surface \cite{moreira2013tutorial}, providing a unique advantage in all weather conditions due to its ability to penetrate clouds and other atmospheric obstructions. Widely employed to monitor various activities, SAR plays a crucial role in tracking urban development \cite{8552453}, assessing biomass changes \cite{chang2022application}, and detecting ships \cite{8124934}, for example. In recent years, the accessibility of SAR data has increased substantially. The availability of numerous datasets has opened up diverse possibilities for applications using neural networks, which often demand substantial data quatity for optimal efficiency \cite{max_aae,9508501}. Among these applications, classification is a crucial task. However, a common hurdle lies in the requirement for labeled images to train neural networks, as most deep learning frameworks rely heavily on it. Addressing this challenge is particularly pertinent in SAR classification, where the Moving and Stationary Target Acquisition and Recognition (MSTAR) dataset stands out as one of the few labeled datasets, extensively employed for benchmarking classification algorithms. \\

This article addresses the classification task in SAR imagery, focusing on overcoming the limitations posed by the scarcity of labeled images. Unlike conventional methodologies that rely predominantly on training and testing algorithms on the same dataset \cite{geng2023target,pei2023self}, we propose a novel approach. Our method involves training a SAR feature extractor (SFE) model based on Vision Transformers (Vit) \cite{dosovitskiy2020image} and contrastive learning \cite{balestriero2023cookbook} on a specific dataset. The goal is to see if it can extract meaningful features from another dataset than the one on which it has been trained, which can lead to good classification performances without fine-tuning. In this case, the training dataset consists of images obtained with the sensor SETHI \cite{9078973} from ONERA, and the test dataset is the MSTAR. In particular, our approach demonstrates excellent performance quantitatively with classification accuracy and qualitatively when features are displayed in a 2D space with a t-SNE algorithm. For classification, it outperforms a ResNet-34 explicitly trained for the task by a large margin in the case of few-shot learning \cite{wang2020generalizing}.\footnote{The code for this article is available at \url{https://github.com/muzmax/MSTAR_feature_extraction.git}}

\begin{figure*}[t]
    \centering
    \includegraphics[width=0.87\linewidth]{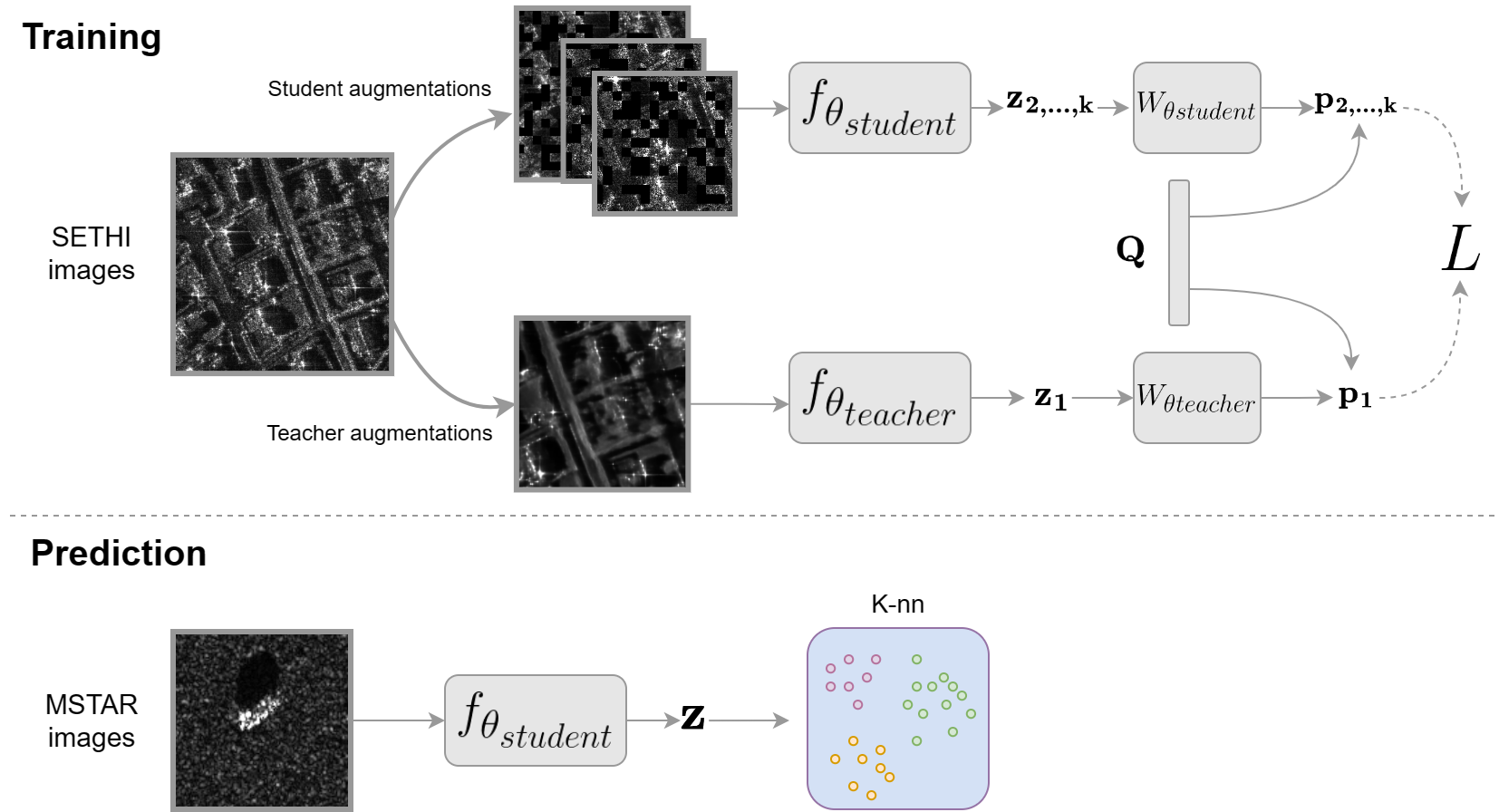}  
    \caption{SAR feature extractor (SFE) architecture for the training phase and the prediction phase. The notations are detailed in \ref{subsec:arch}. Training images were acquired with the SETHI sensor, and test images are from the MSTAR dataset. The acquisition method, resolution, and dynamic are not the same, and no fine-tuning is performed between training and prediction. }
    \label{fig:archi}
\end{figure*}
\label{sec:intro}

\section{Proposed method}
In this section, we first briefly describe the goal of Siamese networks, the reason ViT is the architecture chosen to extract features from SAR images, the training method and the augmentations used, knowing that there is a significant difference between SAR and optical images. The objective is to train a network on a specific sensor with the following procedure and test the feature extraction efficiency on a different sensor to see how well it generalizes.
\label{sec:method}
\subsection{Model architecture}
\label{subsec:arch}
The methodology used in this study is based on Siamese networks \cite{bromley1993signature}. Two identical networks, commonly referred to as the `student and the teacher',  with different weights, are trained in parallel. In our case, many augmented views are generated from one image, the teacher will encode one augmented image, and the student will encode every other. The objective is to train the student network to produce encoded feature vectors that show a high similarity to the vector generated by the teacher network. \\

The architecture is based on the latest advances in self-supervised learning for computer vision \cite{caron2021emerging,assran2022masked}. These methods are convenient for Earth observation for several reasons. The training only needs positive pairs of images, which is simpler to define compared to negative pairs. One other advantage is that the architecture is based on ViT's \cite{dosovitskiy2020image}. Unlike traditional convolutional neural networks, such as CNNs, ViTs process images as sequences of patches and leverage self-attention mechanisms to capture global dependencies. One of their key features is their ability to handle input images of different shapes without the need for resizing. An image $\mathbf{X}\in \mathbb{R}^{h\times w\times c}$ is encoded in $p$ patches of fixed size $d_e$ with a convolutional layer, such that $\mathbf{X}_{enc} = [\mathbf{x}_1,\mathbf{x}_2,\ldots,\mathbf{x}_p]\in \mathbb{R}^{d_e\times p }$. A positional encoding is added for each $\mathbf{x}_i$ to add spatial context, and then $\mathbf{X}_{enc}$ is multiplied by multiple weight matrices to obtain the embedding, as explained in \cite{dosovitskiy2020image}. For a given weight matrix $\mathbf{W}_e$, we have the projection $\mathbf{W}_e \, \mathbf{X}_{enc}$. Since the linear projection is applied element-wise to each patch vector, the model can naturally handle varying image sizes. Because SAR resolution can significantly vary between sensor types and also because we might not want to extract features for the same application, it is convenient to have the possibility to input images of any size. The use of ViTs also allows us to use a specific augmentation that relies on the encoded patches, as we will see later.\\

We use the loss function and regularizations detailed in \cite{assran2022masked} to train the network. Each image is encoded by a ViT ($f_{\theta_{student}}$ or $f_{\theta_{teacher}}$) in a feature $\mathbf{z}\in \mathbb{R}^{d_e}$, then projected with a multilayer perceptron head ($\mathbf{W}_{\theta _{student}}$ or $\mathbf{W}_{\theta_{teacher}} $) in a representation $\mathbf{h}\in \mathbb{R}^{d_h}$. The result is then projected onto a set of prototypes $\mathbf{Q} = [\mathbf{q}_1,\mathbf{q}_2,\ldots,\mathbf{q}_n]\in \mathbb{R}^{d_h \times n}$ and $\mathbf{s} = (\mathbf{s}_1,\mathbf{s}_2,\ldots,\mathbf{s}_n)^T \in \mathbb{R}^{n}$      such that 
\begin{equation}
    \mathbf{p} = \mathrm{softmax}\left(\displaystyle\frac{\mathbf{s}}{\tau}\right)  \text{ with } \left\{\mathbf{s}_i = \displaystyle\frac{\mathbf{q}_i^T \mathbf{h}}{\left\| \mathbf{q}_i\right\|_2 \, \left\| \mathbf{h}\right\|_2}\right\}_{i\in [1,n]}\, , \label{eq:proba}
    \end{equation}   
and where $\tau$ is a temperature to soften or sharpen the distribution (we set $\tau$ smaller for the teacher than for the student to force the network to have a sharper prediction). The vector $\mathbf{p}$ is the softmax of the cosine similarity between the image representation $\mathbf{h}$ and each prototype. Having this value instead of simply $\mathbf{h}$ forces the network to encode similar images in the same cluster. This method proves to be advantageous for applications such as target classification, where the objective requires a distinct clustering of features.\\

For a given batch of size $b$, a total of $k$ augmentations are computed for each image. As we will see in Section \ref{subsec:data}, the student and teacher augmentations differ. The first one will be given to the teacher and the rest to the student. The training loss will be decomposed into a similarity loss:
\begin{equation}
L_{sim} = \frac{1}{b~(k-1)} \sum \limits_{i = 1}^b \sum \limits_{j = 2}^k \sum_{l=1}^n  -\mathbf{p}_{i,1}^l~\log{\mathbf{p}_{i,j}^l}\, ,
\end{equation}
and an entropy maximization regularizer $R$ to ensure that every prototype will be used to cluster the data:
\begin{equation}
R = -\sum_{l=1}^{n}\overline{\mathbf{p}}^l ~\log~ \overline{\mathbf{p}}^l  \text{ with } \overline{\mathbf{p}} = \frac{1}{b~(k-1)} \sum \limits_{i = 1}^b \sum \limits_{j = 2}^k \mathbf{p}_{i,j} \, ,  
\end{equation}
where $\mathbf{p}_{i,j}^l$ describes the l-component of the $n$-vector computed in \eqref{eq:proba} for j-th augmented view of the i-th image of a batch and $\bar{\mathbf{p}}^l$ denotes the l-component of the $n$-vector $\bar{\mathbf{p}}$. The final loss is given by $L = L_{sim} - \lambda \, R$, where $\lambda$ is a positive number used to weight the importance of $R$. The student network is the only one updated with backpropagation, the teacher is updated with a moving average such that $\theta_{teacher} \leftarrow m\, \theta_{teacher} + (1-m)\,  \theta_{student}$.

\subsection{Data augmentation}
\label{subsec:data}
The augmentation method is crucial because it will tell the network what images should be encoded with similar features. Many standard data augmentation methods used in contrastive learning, such as color distortion, rotation, and blur \cite{chen2020simple}, are not suitable for SAR imaging. 

Fortunately, there is one augmentation technique that yields a significant performance improvement and is adaptable for SAR images, which is masking parts of the input image. Two masking strategies are employed; one is a basic rectangular crop without resizing. It can be local or global (small or large window). The second method is a random mask of input patches $[\mathbf{x}_1,\mathbf{x}_2,\ldots,\mathbf{x}_p]$. In addition to being practical and easily applicable augmentations, it scales well with the dataset size by reducing training time and memory requirements. 

Instead of using a blur, a subsampling augmentation based on SAR subband extraction is used \cite{Brekke2013}. The SLC image spectrum is cropped in its center before returning to the spatial domain. 

Before entering the network, a log transformation normalized between 0 and 1 is applied to reduce the image's dynamic range. The data mean value is then randomly shifted to mimic the optical color distortions.

SAR images naturally contain a strong perturbation called speckle \cite{goodman1976some}. Instead of adding noise as an augmentation, we use a despeckling network to remove these fluctuations. In this case, the MERLIN architecture proposed in \cite{dalsasso2021if} is used and trained on our dataset.

The augmentations experienced for the image sent to the teacher network are composed of one global crop in addition to the despeckling process. As for the student, there are one global and multiple local crops, a subsampling, and a mean shift used for the augmentations.

\subsection{Performances assesment}
\label{subsec:perf}
Only the student ViT will be kept for a qualitative and quantitative evaluation. Then, it will extract the features of the targets from a sensor that it has never seen. To see if the representation is pertinent, a k-NN will be carried out for classification, and the features will be projected on a 2D space. A summary of the method is explained in Fig.\ref{fig:archi}.

\section{Experiments}
First, this section explains the training data and model parameters. Then, the model is evaluated with qualitative and quantitative tests. For the quantitative part, we compare the results of the trained network with a ResNet-34 architecture trained on the MSTAR dataset and with a k-NN applied on the data reduced with PCA. This is done in the specific case of few-shot learning.

\subsection{Training}
The X-band and L-band training images are acquired by SETHI, the airborne SAR developed by ONERA \cite{9078973}. They have a resolution of 20cm and 1m in both azimuth and range domains. In total, there are 199040 patches of size $100\times100$ pixels. The global and local crop sizes are $64\times64$ and $32\times32$, respectively. For the student, we create three local crops and one global. The network is a tiny ViT architecture with a patch size of 8 (ViT-T/8) and 256 prototypes, trained for 600 epochs with the hyperparameters described in \cite{assran2022masked}.

\subsection{Evaluation dataset}
The proposed network will then be tested on the MSTAR dataset. It is composed of 30 cm X-band images of seven different types of vehicles and a calibration class. These images are pretty different from SETHI images. They are also X-band images but were acquired in stripmap mode for SETHI and in spotlight mode for the MSTAR dataset. The two data are not normalized similarly, so the dynamic also differs. And finally, the image sizes are different. In training, patches are of sizes $64\times64$, $32\times32$, and $16\times 16$ with sub-resolution augmentation. Whereas with the MSTAR dataset, images can have, for example, sizes of $128\times 128$, $54\times 54$, and $192\times 192$. The number of images per class is summarized in Table.~\ref{table:res}.
\begin{table}[!h]
\begin{center}
\begin{tabular}{|c |c| c| c| c|} 
 \multicolumn{5}{c}{MSTAR dataset} \\ [0.5ex] 
 \hline 
 Class & 2S1 & BRDM\_2 & BTR\_60 & D7 \\ [0.5ex] 
 \hline
 Number & 1664 & 1282 & 451 & 573 \\ [1ex] 
 \hline
 Class & T62 & ZIL131 & ZSU\_23\_4 & SLICY \\ [0.5ex] 
 \hline
 Number & 572 & 573 & 1401 & 2539 \\ [1ex] 
 \hline
\end{tabular}
\caption{\label{table:res}Image number for each class. The calibration class is called "SLICY" and the others are vehicles.}
\end{center}
\end{table}

\begin{figure}[t]
    \centering
    \includegraphics[width=\linewidth]{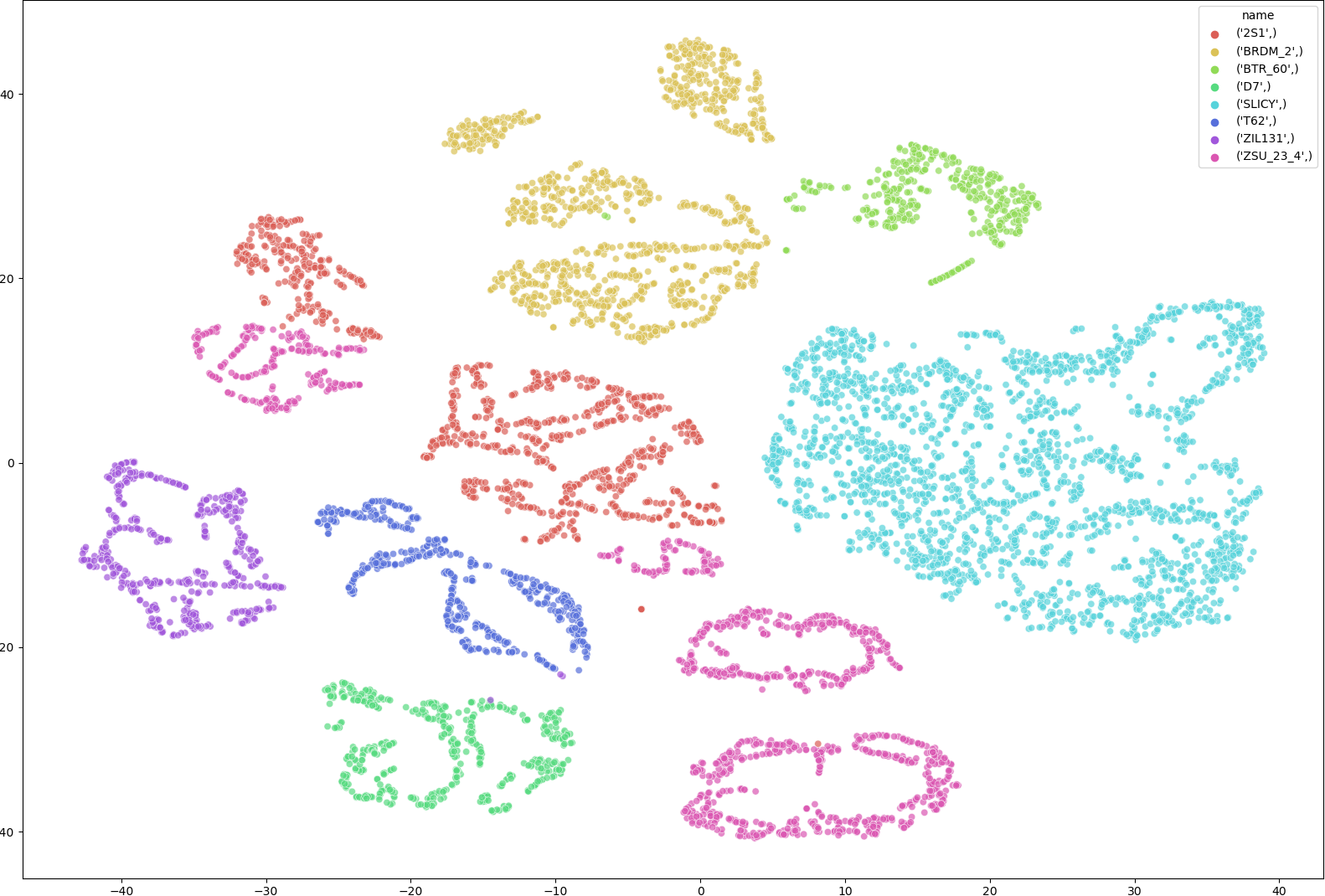}  
    \caption{t-SNE 2D projection of MSTAR encoded images.}
    \label{fig:tsne}
\end{figure}
\subsection{Feature visualization}
    To qualitatively assess the relevance of the extracted features, we present a 2D visualization in Fig.~\ref{fig:tsne} generated using the t-SNE algorithm, in which each image is encoded in a vector $\mathbf{z}$ (see Fig.~\ref{fig:archi}) and then compressed in a 2D vector. Notably, despite the network not being trained on MSTAR data or any military vehicle, its clustering capabilities show promise. Some classes, such as "SLICY" or "ZIL131," are separated from others. Conversely, for classes like "2S1" and "ZSU\_23\_4"  the clustering results in multiple distinct subclusters. This visual representation correlates directly with the performance of the k-NN algorithm. Specifically, if a class is perfectly clustered, a single image from that class would be sufficient to classify the entire dataset accurately.

\subsection{Classification performances}
A quantitative evaluation follows the procedure explained in \ref{subsec:perf}. The k-NN classification of the extracted features is evaluated in the case of few-shot learning.  For each class of the MSTAR dataset, the number of labeled images varies between one and a hundred, with an emphasis between one and ten. To assess the performance of this method, a ResNet-34 is trained with the same number of labeled images to perform the classification. Conjointly, a PCA is done following the same procedure as it is explained in Fig\ref{fig:archi} for the prediction phase, except that the PCA is used as the feature extractor. Both the ResNet and PCA have a pre-processing step with a log transformation followed by a normalization between 0 and 1 and a resizing. The resizing dimensions are set at $224\times 224$ for the ResNet and $150\times 150$ for the PCA. Each ResNet is trained for 200 epochs and the evaluation dataset includes all the data except the hundred images per class used in training.

\begin{figure}[t]
    \centering
    \includegraphics[width=\linewidth]{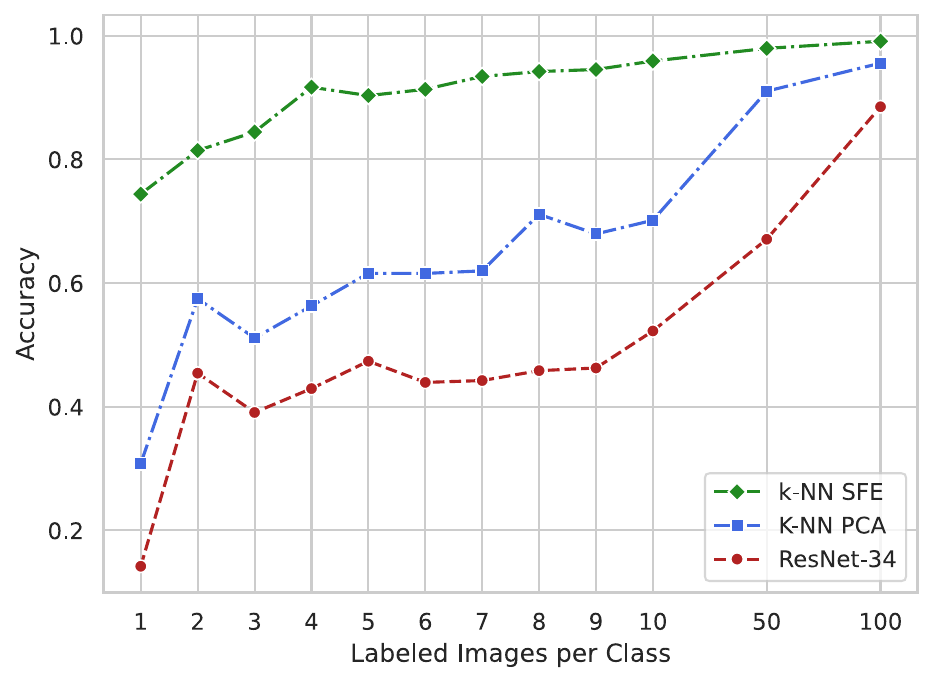}  
    \caption{Comparison of few-shot classification accuracy, where 'k-NN SFE', 'k-NN PCA' and 'ResNet-34' represent respectively our method, a k-NN on the data reduced with PCA and A ResNet-34 trained from scratch.}
    \label{fig:acc}
\end{figure}

As illustrated in Fig.~\ref{fig:acc}, the performance obtained using a k-NN (with k=2) classifier on the SFE outperforms both the convolution network and the k-NN with PCA. With only one labeled image per class, our method attains a 74\% accuracy, which rises to 95.9\% with ten labeled images. This marks a difference of 43.7\% compared to the ResNet and 25.9\% compared to the k-NN with PCA, achieving respective accuracies of 52.2\% and 70.16\% with ten labeled images per class. Contrary to our method, a standard convolutional network requires more labeled images to achieve good results. Even when the image count increases to a hundred, the performance gap between the methods narrows, but the SFE maintains superior accuracy.
\section{Conclusion}
In this paper, we propose a new deep-learning framework based on contrastive learning for SAR feature extraction in the case of classification. It underscores great adaptability across sensor types. Although the feature extractor has not seen a single image of the MSTAR dataset, it can cluster each class accurately. This method leads to great classification performances when a k-NN algorithm is used on top of the extracted features. It outperforms the accuracy of a ResNet-34 in the case of few-shot learning. Even when the number of labeled data increases, the accuracy of the k-NN remains better, reaching 99.1\% with 100 labeled images per class. The proposed method shows great promise in establishing a versatile feature extractor model for SAR images, applicable across various sensors and diverse applications.

\bibliographystyle{IEEEbib}
\bibliography{strings,refs}

\end{document}